\documentclass[letterpaper,10pt,conference]{ieeeconf}
\pdfoutput=1

\IEEEoverridecommandlockouts                              

\usepackage{graphics} 
\usepackage{epsfig} 
\usepackage{mathptmx} 
\usepackage{times} 
\usepackage{cite}
\usepackage{amsmath,amssymb,amsfonts}
\usepackage{algorithmic}
\usepackage{graphicx}
\usepackage{textcomp}
\usepackage{xcolor}
\usepackage[hyphens]{url}

\usepackage{hyperref}
\hypersetup{hidelinks}
\usepackage{cleveref}
\usepackage{algorithm}
\usepackage{gensymb}
\usepackage{booktabs}
\usepackage{float}
\usepackage{graphicx}
\usepackage{bm}
\usepackage{amsfonts}
\usepackage{stmaryrd}
\usepackage{cancel}
\usepackage{mathtools}

\usepackage{subcaption}

\usepackage{caption}

\title{Local retail electricity markets for distribution grid services}

\author{Vineet Jagadeesan Nair$^{1}$ and Anuradha Annaswamy$^{1}$
\thanks{*This work was supported by US Department of Energy under Award DOE-OE0000920, the Martin Society Fellowship for Sustainability from the MIT Environmental Solutions Initiative, and a summer internship at the National Renewable Energy Laboratory.}
\thanks{$^{1}$Vineet Jagadeesan Nair and Anuradha Annaswamy are with the Department of Mechanical Engineering, Massachusetts Institute of Technology, Cambridge, MA 02139, USA. {\tt\small \{jvineet9, aanna\}@mit.edu}}}

\begin{document}

\maketitle
\thispagestyle{empty}
\pagestyle{empty}

\begin{abstract}
We propose a hierarchical local electricity market (LEM) at the primary and secondary feeder levels in a distribution grid, to optimally coordinate and schedule distributed energy resources (DER) and provide valuable grid services like voltage control. At the primary level, we use a current injection-based model that is valid for both radial and meshed, balanced and unbalanced, multi-phase systems. The primary and secondary markets leverage the flexibility offered by DERs to optimize grid operation and maximize social welfare. Numerical simulations on an IEEE-123 bus modified to include DERs, show that the LEM successfully achieves voltage control and reduces overall network costs, while also allowing us to decompose the price and value associated with different grid services so as to accurately compensate DERs.

\end{abstract}

\section{Introduction and Motivation}

With increasing penetration of DERs such as renewables, storage, and flexible loads in the distribution system, it is critical to design market structures that enable their smooth integration at the grid edge - to balance variable supply and demand, and increase the utilization of clean, renewable energy sources while maintaining affordability, reliability, and resilience. Local electricity markets (LEMs) have the potential to empower the consumer to take control of their energy footprint, allow transactive energy trading among members of a community, improve community resilience against wider grid events, and potentially reduce energy bills. This paper focuses on one such market, a local electricity market (LEM) at the retail level. This market is designed for DER-rich distribution systems and includes a hierarchical structure, so as to facilitate DER integration, increase market participation of customers and prosumers, and utilize their assets to provide valuable grid services. 

Local energy markets have the capability to allow electricity prices to be endogenous quantities rather than be imposed exogenously. In such a marketplace, prosumers can buy and sell energy in an open marketplace, or through an operator \cite{tohidi2018review}. Normal practice adopted by grid operators and utilities is to rely on standard load profiles from historical data. This is challenged because of the intermittent and highly variable nature of the generation from solar photovoltaic (PV) panels, demand from electric vehicles (EV), and needs of other DERs, as they can cause unpredictable swings in demand and/or generation. LEMs have the potential to help solve this problem for energy retailers and other grid management entities by offering flexibility services and creating opportunities for new business models. The LEM we propose has a two-tier structure, with a Primary Market (PM) at the upper level and a Secondary Market (SM) at the lower level. The SM consists of DER-coordinated assets (DCA) located at each secondary feeder bidding into the market. These DCAs could consist of rooftop PVs, battery storage, or flexible loads. They could have multiple independent owners and thus may not allow a single agent to represent all of them in an aggregated manner. An SM operator (SMO) oversees market operations at this level to clear and schedule DCAs. At the upper level, the SMO has a dual role as an agent in the PM as they are at a node in the primary feeder. All of these SMO agents bid into the PM. These are in turn cleared and scheduled by a PM Operator (PMO), who represents an entire primary feeder.

In our earlier work, we showed the market bidding and clearing process for such an LEM \cite{nair2022hierarchical}. The distribution grid therein, however, was simple, radial, and balanced. In this paper, we relax these assumptions to model unbalanced, multi-phase, and meshed networks using a Current-Injection based linear model for solving AC Optimal Power Flow (ACOPF) at the primary level, employing McCormick envelopes convex relaxations. More importantly, we show in this paper that our LEM can provide valuable services or Volt-VAR control (VVC) and voltage regulation in an unbalanced distributed grid. A distributed Proximal Atomic Coordination algorithm is used for PM clearing which preserves privacy, reduces communication requirements, and improves computational tractability. We also introduce 3-phase pricing at both the SM and PM, to motivate how we can determine the value of such grid services in real-time energy markets based on an optimization framework. We disaggregate the distribution locational marginal prices (dLMP) and local retail tariffs among different SMOs and DCAs, respectively, and decompose their components arising from economic objectives like maximizing social welfare and minimizing costs versus grid objectives like minimizing line losses and voltage profile deviations, while satisfying several power flow constraints. We show that the resulting LEM leads to effective VVC, along with efficient pricing and market compensation through spatial-temporal price differentiation.

\section{Background and Contributions}

\subsection{Prior work}

Several works in the literature have proposed local energy markets for the future power grid with high DER penetration \cite{hvelplund2006renewable,chen2018next,lezema2019lem,lembook} and transactive energy systems more broadly. However, these have not comprehensively studied the potential for such retail electricity markets to provide distribution grid services. Prior work on market design for grid services has largely focused on transmission, for ancillary services like frequency regulation. Some have considered the possibility of active distribution systems to provide services like voltage support to the transmission grid \cite{karagiannopoulos2020active} or the use of reactive power (VAR) assets for voltage balancing and regulation \cite{arnold2016optimal}. However, they have not studied how these services interact with market structures, or how they would be priced. Accurate pricing mechanisms and incentives are critical to coordinate distributed agents and prosumers, where centralized dispatch or direct control is not possible.

While some recent works have explored distribution-level pricing, they have focused either solely on P \cite{haider2020toward,holmberg2022comparison,haider2021reinventing} or notions of Q pricing \cite{potter2023reactive}. A few other works have considered pricing for other grid services like voltage control \cite{bai2017distribution}, but these have relied on using simple power flow models like \textit{LinDistFlow} which ignore line losses, or second-order conic program-based convex relaxations (DistFlow) that are restricted to radial, balanced networks \cite{papavasiliou2017analysis}. These approximations are generally not valid for low to medium-voltage distribution systems which have significant losses and are generally unbalanced and multi-phase. 

\subsection{Our contributions}

We propose the use of our novel LEM architecture for providing grid services, specifically voltage regulation (or Volt-VAR control) in distribution grids. We build upon our prior work that proposed the hierarchical LEM structure consisting of the SM and PM \cite{nair2022hierarchical}, where we used a nonlinear DistFlow (branch flow) model for solving ACOPF at the primary level based on a second-order conic convex relaxation \cite{haider2020toward}. However, this was restricted to radial and balanced systems. We now extend this market structure to unbalanced, multi-phase and meshed networks by instead using a Current-Injection based linear model for ACOPF and PM clearing \cite{ferro2020distributed}. We also introduce notions of three-phase pricing at both the SM and PM levels, and motivate how we can determine the value of such grid services in real-time energy markets based on an optimization framework, in order to accurately compensate DERs and prosumers. This also extends prior work on hierarchical hybrid Volt-VAR control \cite{haider2020toward} by coordinating DERs through market mechanisms and price-based transactive control rather than via direct control of these devices by the grid operator. Finally, we disaggregate the distribution locational marginal prices (dLMP) and local retail tariffs among different SMOs and DCAs, respectively, and break down their components arising from economic objectives like maximizing social welfare versus grid objectives like minimizing line losses and voltage profile deviations.

\begin{figure}
\centering
     \includegraphics[width=\columnwidth]{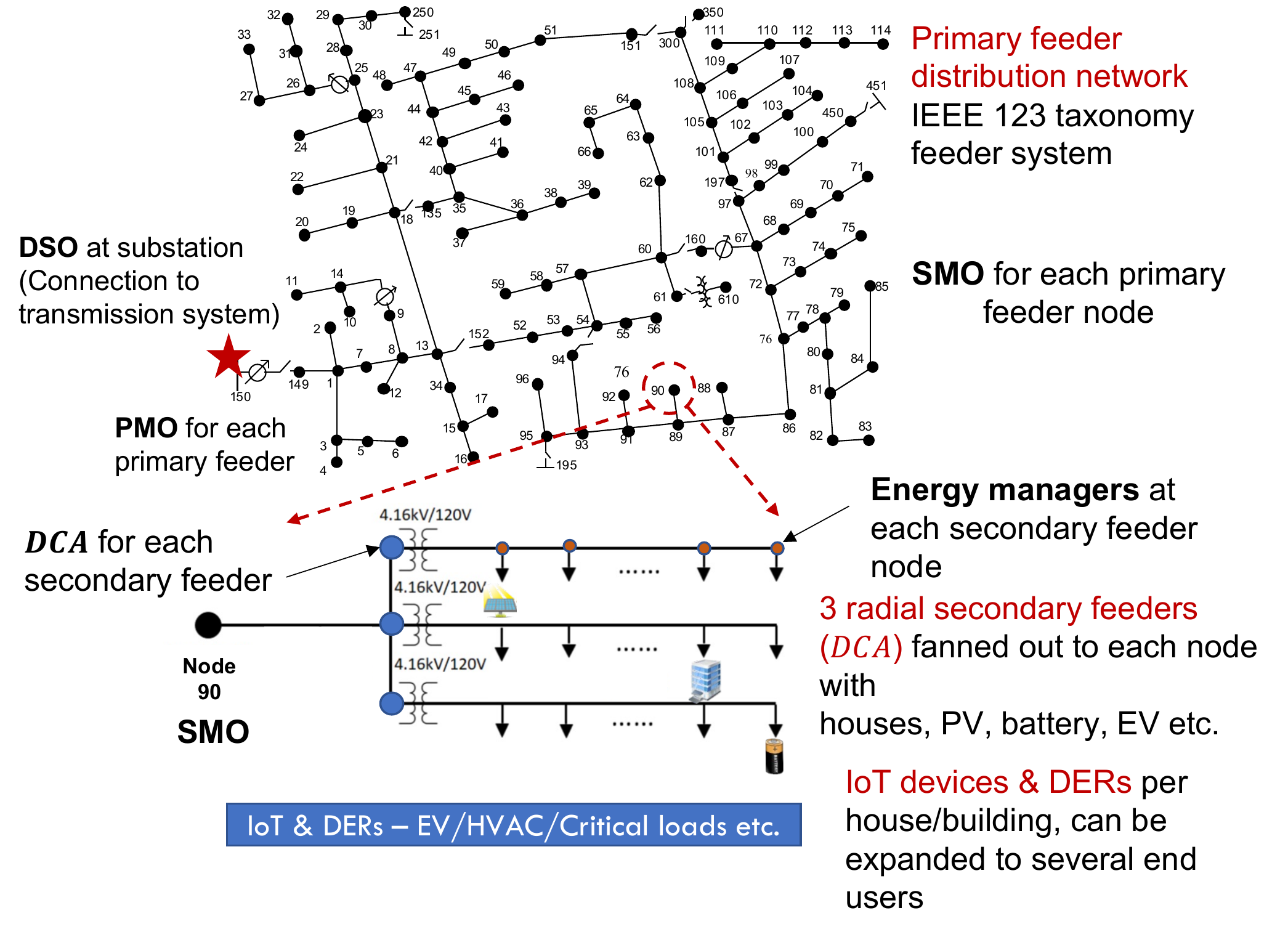}
     \caption{Hierarchical LEM co-located with distribution grid. \label{fig:network_diagram}}
\end{figure}






\section{Methodology}

\begin{figure}
 \centering
     \begin{subfigure}[t]{0.39\columnwidth}
         \includegraphics[width=\linewidth]{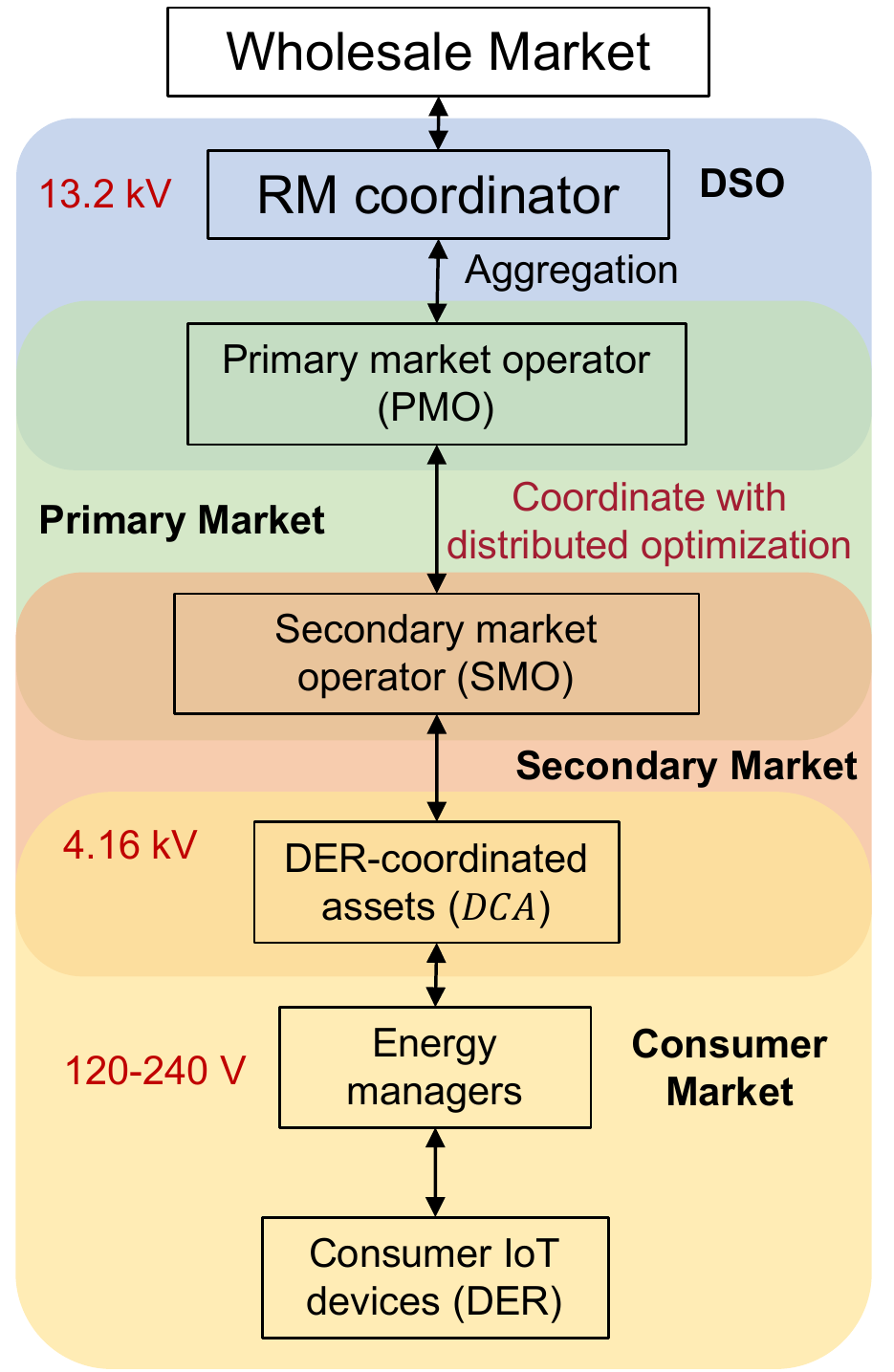}
         \caption{LEM agents. \label{fig:agents}}
     \end{subfigure}
     \begin{subfigure}[t]{0.59\columnwidth}
         \includegraphics[width=\linewidth]{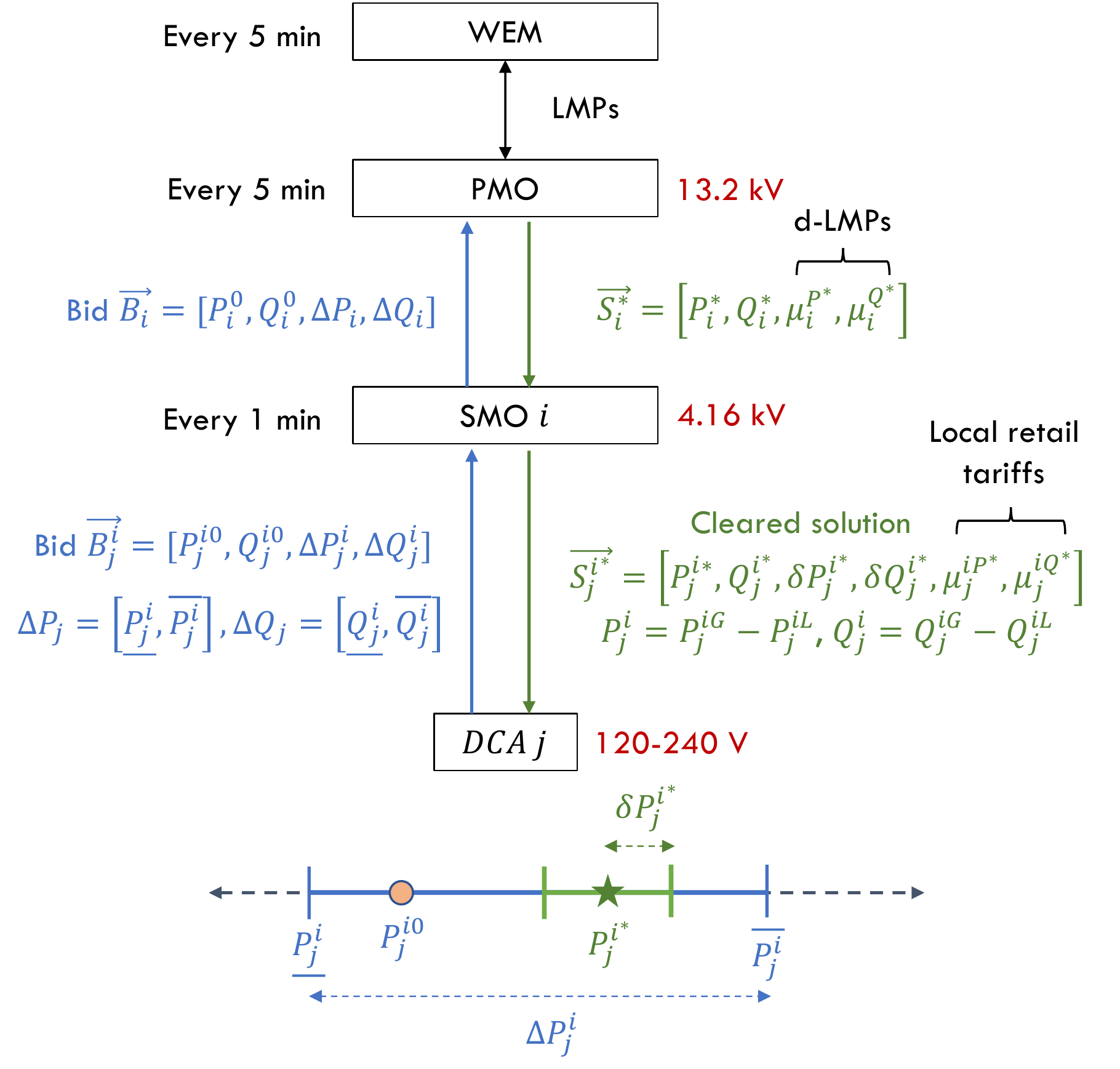}
         \caption{LEM bids and solutions.\label{fig:flowchart}}
     \end{subfigure}
     \caption{Overall structure of the hierarchical LEM.\label{fig:flowchart_agents}}
\end{figure}

We use the LEM structure proposed in our previous work \cite{nair2022hierarchical}, with an SM at the lower level and a PM at the  upper level, as in in \cref{fig:flowchart_agents}. These markets are operated by an SMO and PMO, respectively, with combined oversight of both by a DSO. The upper-level PMO coordinates with the wholesale energy market (WEM), while the lower-level SMO oversees the DCAs. Both the SM and PM clearing use flexibility bids submitted by the DCA and SMO, respectively. Fig. \ref{fig:flowchart} illustrates flexibility bids from a DCA and SMO, in the SM and PM respectively. The SM clearing results in a revised flexibility range for each DCA while the PM clearing results in setpoints for each SMO. These solutions are used to set bilateral contracts between SMOs and DCAs (in the SM), and the PMO and SMOs (in the PM). The following quantities are used to define our SM optimization. For simplicity, we have ignored subscripts and superscripts that indicate the SMO and DCA being considered. For e.g., $P^{i,\phi}_j$ corresponds to DCA $j$ under SMO $i$ while $P_{i,\phi}$ corresponds to SMO $i$, for phase $\phi$.
\begin{itemize}
        \item $j \in \mathcal{N}_{J,i}$: Set of indices of all DCAs under SMO $i$. 
        \item $P^0, Q^0$: Baseline active and reactive power injections.
        \item $\Delta P = [\underline{P},\overline{P}$, $\Delta Q = [\underline{Q},\overline{Q}]$: Bid flexibilities for each DCA, giving the range of maximum downward and upward flexibilities in $P$ and $Q$ injections offered by the DCA.
        \item $t_p$ and $t_s$: Timestamps for the PM and SM, respectively.
        \item $\Delta t_p, \Delta t_s$ and $\Delta t_{WEM}$: Time periods for the PM, SM, and WEM, respectively. Here, we assumed that the SM is cleared more frequently, every $\Delta t_s$ = 1 minute, while the PM is cleared every $\Delta t_p$ = 5 minutes.
        \item $\hat t_p$: PM clearing prior to current SM interval [$t_s,t_s+\Delta t_s$].
        \item $P^*(\hat{t}_p), \; Q^*(\hat{t}_p)$: Setpoints provided by the PM to SMOs.
\end{itemize}
\subsection{Secondary market model}
The SMO $i$ solves the multiobjective optimization problem shown in \cref{eq:SMopt} to schedule the DCAs $j$ based on their flexibility bids. We model both net load and net generator DCAs at the secondary feeder level separately, indicated by superscripts $K \in \{L,G\}$, respectively. All power injections are three-phase (and possibly unbalanced) variables for each DCA or secondary feeder. Terms without a superscript $K$, $G$ or $L$ refer to net injections (generation minus load), i.e., $P^i_j = P^{iG}_j - P^{iL}_j$. All variables are specified for the current secondary timestep $t_s$ unless stated otherwise. 
\begin{subequations}
\label{eq:SMopt}
\begin{align}
& \min \sum_{j \in \mathcal{N}_{J,i}} \{f_{j,1}^i,f_{j,2}^i,f_{j,3}^i,f_{j,4}^i\} \label{eq:cost} \\
& f_{1,j}^i \succ f_{2,j}^i \succ f_{3,j}^i \succ f_{4,j}^i \label{eq:ranking}, \; \; K \in \{L,G\}, \; \; \Phi = \{a, b, c\} \\ 
& f_{j,1} = -C^i_j \left(\sum_{\phi \in \Phi}(P_j^{iK,\phi} - P_j^{iK0,\phi})^2 + (Q_j^{iK,\phi} - Q_j^{iK0,\phi})^2 \right)\nonumber \\
& f_{j,2}^i  = \sum_{\phi \in \Phi} \mu_j^{iP,\phi} P_j^{i,\phi} + \mu_j^{iQ,\phi} Q_j^{i,\phi}, \; f_{j,3}^i = -\sum_{\phi \in \Phi}(\delta P_j^{iK,\phi} + \delta Q_j^{iK,\phi}) \nonumber\\
& f_{j,4}^i = \beta_j^{iP}\sum_{\phi \in \Phi}\left(P_j^{iK} - P_j^{iK0}\right)^2 + \beta_j^{iQ}\sum_{\phi \in \Phi}\left(Q_j^{iK} - Q_j^{iK0}\right)^2 \nonumber \\
& \text{subject to:} \nonumber\\
& P_j^{iK,\phi} - \delta P_j^{iK,\phi} \geq \underline{P}_j^{iK,\phi} \; Q_j^{iK,\phi} - \delta Q_j^{iK,\phi} \geq \underline{Q}_j^{iK,\phi}  \label{eq:PQ_lowerlim} \\
& P_j^{iK,\phi} + \delta P_j^{iK,\phi} \leq \overline{P}_j^{iK,\phi}, \; Q_j^{iK,\phi} + \delta Q_j^{iK,\phi} \leq \overline{Q}_j^{iK,\phi}  \label{eq:PQ_upperlim} \\ 
& \delta P_j^{iK,\phi}, \; \delta Q_j^{iK,\phi} \geq 0, \; 0 \leq \mu_j^{iP} \leq \overline{\mu}^{iP}, 0 \leq \mu_j^{iP} \leq \overline{\mu}^{iQ}  \; \label{eq:flex_fairness} \\ 
& \sum_{t_s}^{t_s + \Delta t_p} \sum_{j \in \mathcal{N}_{J,i}} \sum_{\phi \in \Phi} \left(\mu_j^{iP,\phi}(t) P_j^{i,\phi}(t) + \mu_j^{iQ,\phi}(t) Q_j^{i,\phi}(t)\right) \Delta t_s  \nonumber \\
& \leq \sum_{\phi \in \Phi} \left(\mu_i^{P^*,\phi}(\hat{t}_p) P_i^{\phi^*}(\hat{t}_p) + \mu_i^{Q^*,\phi}(\hat{t}_p) Q_i^{\phi^*}(\hat{t}_p)\right) \Delta t_p \label{eq:budgetPQ} \\
& \sum_{j \in \mathcal{N}_{J,i}} P_j^{i,\phi}(t_s) = P_i^{\phi^*}(\hat{t}_p), \quad \sum_{j \in \mathcal{N}_{J,i}} Q_j^{i,\phi}(t_s) = Q_i^{\phi^*}(\hat{t}_p) \label{eq:PQbalance}
\end{align}
\end{subequations}

The following decision variables are determined as outputs of the optimization for each DCA $j$ bidding to SMO $i$, determined at $t_s$ and applied over the period $[t_s, t_s+\Delta t_s]$: 
\begin{itemize}
    \item $P_j^{i,\phi},Q_j^{i,\phi}$: Optimal power injection setpoints.
    \item $[\delta P_j^{i,\phi}, \delta Q_j^{i,\phi}]$: Optimal symmetric flexibility ranges around setpoints, i.e. DCAs are directed to have net injections within these intervals $[P_j^{i,\phi} - \delta P_j^{i,\phi}, P_j^{i,\phi} + \delta P_j^{i,\phi}]$.
    \item $\mu_j^{iP},\mu_j^{iQ}$: Local electricity tariffs.
\end{itemize}
The commitment score $C_j^i(t) \in [0,1]$ reflects the SMO's confidence in whether the DCA $j$ will reliably follow their committed injections within the flexibility range specified above, with higher values $C_j^i$ indicating more reliable assets. This is updated based on the deviations of actual DCA responses ($\hat{P}_j^i$) from their scheduled setpoints ($P_j^{i^*}$) and ranges ($[\underline{P}_j^{i^*},\overline{P}_j^{i^*}]$) in the SM (see \cite{nair2022hierarchical} for details).

The cost functions in~\eqref{eq:SMopt} correspond to the following: 
\begin{enumerate}
    \item \textbf{Commitment reliability $\mathbf{f^i_{j,1}}$}: Maximize flexibility of injections assigned to more trustworthy DCAs (i.e., $C^i_j$ closer to 1) while minimizing the flexible scheduling of DERs with lower commitment scores, who are less likely to abide by their contractual commitments.
    \item \textbf{Net costs $\mathbf{f^i_{j,2}}$}: Minimize \textit{net} costs to the SMO for running its SM, comprising payments from the SMO to DCAs that are net generators, and denote revenue from the DCAs that are loads.
    \item \textbf{Flexibility $\mathbf{f^i_{j,3}}$}: Maximize aggregate flexibility that the SMO can extract from its DCAs, and offer to the PMO.
    \item \textbf{Disutility $\mathbf{f^i_{j,4}}$}: Minimize inconvenience to DCAs when they provide flexibility to the operator. Thus, our SMO is an altruistic entity that also considers welfare maximization for its DCAs. For our simulations, disutility coefficients were $\beta^{iP}_j, \beta^{iQ}_j \sim \mathcal{U}[0.1,1]$.
\end{enumerate}

The constraints in \cref{eq:PQ_lowerlim,eq:PQ_upperlim} reflect the feasible power injections for each of the DCAs, which are determined by their capacity limits and flexibility bids. Note that while generators can submit both upward and downward flexibilities from their baseline injection values, loads can only offer only downward flexibility, i.e. $\overline{P}_j^{iL} = P_j^{i0}$. 
The budget constraint \cref{eq:budgetPQ} ensures that the total payments by the SMO to its DCAs for all SM clearings within each primary interval are less that its net revenue received from the PMO during the same period, in order to remain solvent. Notice that all the terms in the objective function are either linear or convex except for the net bilinear cost term $f^i_{j,2}$ which is not convex in general. Similarly, all the constraints are linear except for the bilinear budget balance inequality constraint \cref{eq:budgetPQ}. In our prior work \cite{nair2022hierarchical}, we solved the full nonconvex optimization problem in order to obtain retail tariffs directly as outputs of the SM clearing. However, in this work we consider a simpler, convexified version of the problem. We achieve this by removing the budget constraint \cref{eq:budgetPQ} from the original problem and instead enforcing exact budget balance ex-post by setting the retail tariffs for each of the DCAs after the fact. This also implies that the SMO is no longer explicitly minimizing costs in the SM (i.e. we remove the $f^i_{j,2}$ term from the objective) and is instead just breaking even, essentially acting as a purely non-profit entity and market maker. By transforming the inequality constraint in \cref{eq:budgetPQ} to a strict equality, the SMO can derive the following localized real-time retail tariffs for each DCA $j$ at every secondary timestep $t_s$ using their cleared $P$ and $Q$ schedules:
\begin{align}
\mu^{iP}_j(t_s) & = y^{iP}_j(t_s) \frac{|R_{PM}|}{|P_j^{i^*}(t_s)|\Delta t_s}, \; \mu^{iQ}_j(t_s) = y^{iP}_j(t_s) \frac{|R_{PM}|}{|Q_j^{i^*}(t_s)|\Delta t_s} \nonumber \\
R_{PM} & = \left(\mu_i^{P^*}(\hat{t}_p) P_i^*(\hat{t}_p) + \mu_i^{Q^*}(\hat{t}_p) Q_i^*(\hat{t}_p)\right) \Delta t_p \label{eq:breakeven_tariff}
\end{align}
Here $R_{PM}$ is the net revenue to the SMO $i$ from the most recent PM clearing, while $P^*, \; Q^*$ indicate net injections. In order to balance its budget, the SMO sets the price multipliers $y^{iP}_j(t_s), \; y^{iQ}_j(t_s)$ for each of the DCAs and at each time $t_s$ using the following heuristics, which can be derived from the exact budget balance equality constraint. If the DCA $j$ is a net generator in terms of P or Q injections (summed over all its non-zero phases) i.e. $j \in S_G^{P,Q}(t_s) = \{j \in \mathcal{N}_{J,i}: P^{i^*}_j(t_s) > 0 \; \text{or} \; Q^{i^*}_j(t_s) > 0\}$, the price multipliers are:
\begin{equation}
    y^{iP, Q}_j(t_s) = \begin{cases} 0,  & \text{if} \; |S_G^{P,Q}(t_s)| = 0 \\
                                  \frac{1}{2|S_G^{P,Q}(t_s)|} & \text{if} \; |S_L^{P,Q}(t_s)| = 0 \\
                                  1 & \text{otherwise}                               
\end{cases}
\end{equation} 

If DCA $j$ is a net load i.e. $j \in S_L^{P,Q}(t_s) = \{j \in \mathcal{N}_{J,i}: P^{i^*}_j(t_s) < 0 \; \text{or} \; Q^{i^*}_j(t_s) < 0\}$, the price multipliers are:
\begin{equation}
    y^{iP, Q}_j(t_s) = \begin{cases} 0,  & \text{if} \; |S_L^{P,Q}(t_s)| = 0 \\
                                  \frac{1}{2|S_L^{P,Q}(t_s)|} & \text{if} \; |S_G^{P,Q}(t_s)| = 0 \\
                                  \frac{1 + 2|S_G^{P,Q}(t_s)|}{2|S_L^{P,Q}(t_s)|} & \text{otherwise}         \end{cases}
\end{equation}
where $|S^{P,Q}_{G,L}(t_s)|$ denotes the cardinality of the set of DCA generators or loads at time $t_s$. The retail tariffs are assumed to be identical across all phases for each DCA, and are set based on the net power injections summed over all its non-zero phases, i.e., $P^{i^*}_j(t_s) = \sum_{\phi} P^{i,\phi^*}_j(t_s)$. Note that all the above prices are solved for in terms of [\$/kWh] or [\$/kVARh]. Finally, \cref{eq:PQbalance} denote power balance constraints for the SMO, requiring that injections from the DCAs downstream must equal the net flows $P_i^*(\hat{t}_p)$ and $Q_i^*(\hat{t}_p)$ from the primary feeder upstream, as scheduled by the PMO. Since the PM clears less often, these values can be treated as constant for the SM optimization over each $\Delta t_p$.

The multiobjective optimization problem \cref{eq:SMopt} is solved using a hierarchical approach \cite{gunantara2018review,mausser2006normalization} since our objective terms considered in \cref{eq:cost} have different units and may not be comparable in magnitude. The SMO ranks their objectives as shown in \cref{eq:ranking}, setting commitment reliability as their most important goal and disutility as being the least important. The SMO then solves a series of optimization problems, sequentially optimizing these objectives one at a time, in descending order of importance. At each step, constraints are added on how much the previous objective value can be degraded, controlled by the parameter $\epsilon = 5\%$:
\begin{align}
    \min_{\vec{S}_j^i} & \; F_k = \sum_{j \in \mathcal{N}_{J,i}} f_{j,k}^i (\vec{S}_j^{i}) \; \forall \; k = 1, 2, 3 \\
    \text{s.t.} \; & f_{j,\ell}^i(\vec{S}_j^{i}) \leq (1 + \epsilon)\sum_{j \in \mathcal{N}_{J,i}}  f_{j,\ell}^i (\vec{S}_j^{i^*}) = (1 + \epsilon)F_{\ell}^*, \label{eq:degrade} \\
    & \forall \; \ell = 1,2,\dots, k-1, \; k>1, \text{ and constraints} \; \cref{eq:PQ_lowerlim}{\text -}(\ref{eq:PQbalance}) \nonumber
\end{align}

\subsection{Primary market model}

Before each PM clearing, SMO $i$ aggregates schedules of its DCAs $j$ from the last SM clearing, and uses this to bid into the PM. Its bid is similar to the DCA, with baseline injections and flexibilities both determined by SM solutions:
\begin{align}
        P^0_i(t_p) & = \sum_{j \in \mathcal{N}_{J,i}}  P_j^{i^*}(t_p), \; \Delta P_i = \left[\underline{P_i^{\phi}}, \overline{P_i^{\phi}} \right] \nonumber\\
        & \left[\sum_{j\in \mathcal{N}_{J,i}} P_j^{i^*} - \delta P_j^{i^*}, \sum_{j\in \mathcal{N}_{J,i}} P_j^{i^*} + \delta P_j^{i^*} \right]
\end{align} 
The PMO accepts these bids and then clears the PM at every $t_p$ at intervals of $\Delta t_p$. The PMO clears using a dual ascent-based distributed optimization algorithm, known as proximal atomic coordination (PAC) that facilitates market clearing \cite{romvary2021proximal}. Compared to centralized optimization, such a distributed approach helps preserve privacy, reduces computational burden, and improves scalability for large networks since each SMO only needs to locally exchange information with neighboring agents, while still approaching globally optimal solutions. The PM optimization solves the  alternating current (AC) optimal power flow (OPF) problem using a current injection (CI) model as proposed in \cite{ferro2020distributed}. 

\subsubsection{Current injection model}
The primal decision variables for each SMO $i$ obtained by solving the optimization problem $x = [P_i^{\phi},Q_i^{\phi},V_{i}^{\phi,R},V_{i}^{\phi,I},I_{i}^{\phi,R},I_{i}^{\phi,I}]$ consists of (i) active ($P_i^{\phi^*}$) and reactive ($Q_i^{\phi^*}$) power setpoints (ii) real and imaginary components of nodal voltages ($V_{i}^{\phi,R^*}, V_{i}^{\phi,I^*}$) and current injections ($I_{i}^{\phi,R^*}, I_{i}^{\phi,I^*}$). Note that these are solved for each non-zero phase $\phi \in \mathcal{P} = \{a,b,c\}$. The CI-OPF problem formulation is given by:
\begin{subequations}
\label{eq:ciopf}
\begin{align}
&\min _x f^{obj}(x)\\
&I^R=\operatorname{Re}(\mathrm{Y} V), \; I^I=\operatorname{Im}(\mathrm{YV}) \label{eq:ohmslaw} \\
&P_i^\phi=V_i^{\phi, R} I_i^{\phi, R}+V_i^{\phi, I} I_i^{\phi, I} \quad \forall i \in \mathcal{N}, \phi \in \mathcal{P} \label{eq:bilinearP} \\
&Q_i^\phi=-V_i^{\phi, R} I_i^{\phi, I}+V_i^{\phi, I} I_i^{\phi, R} \quad \forall i \in \mathcal{N}, \phi \in \mathcal{P} \label{eq:bilinearQ}\\
& (I_{ij}^{R})^2+(I_{ij}^{I})^2\le \overline{I_{ij}}^2 \quad \forall i\in \mathcal{N}, \phi\in \mathcal{P} \label{eq:Ibranchlims} \\
& \underline{{V_{i}^\phi}}^2\le({V_{i}^{\phi,R}})^2 + ({V_{i}^{\phi,I}})^2 \le \overline{V_{i}^{\phi}}^2 \quad \forall i\in \mathcal{N}, \phi\in \mathcal{P} \label{eq:Vmaglims} \\
& \underline{P_i^\phi} \leq P_i^\phi \leq \overline{P_i^\phi}, \;  \underline{Q_i^\phi} \leq Q_i^\phi \leq \overline{Q_i^\phi}
\end{align}
\end{subequations}
where $Y$ is the 3-phase bus admittance matrix for the network, and $V$ and $I$ are matrices of nodal voltages and currents respectively. Problem \cref{eq:ciopf} is nonconvex due to bilinear constraints \cref{eq:bilinearP,eq:bilinearQ}, and the ring constraint \cref{eq:Vmaglims} on voltage magnitudes. We obtain a convex relaxation by using McCormick envelopes (MCE), which represent the convex hull of a bilinear product $w=xy$ by using upper and lower limits on $x, \; y$. Thus, we replace the bilinear equality with a series of linear inequalities, denoted as $\text{MCE}(w) = \{w=xy: x\in [ \underline{x}, \overline{x}],  y\in [ \underline{y}, \overline{y}]\}$:
\begin{equation} \label{eq:mce}
  MCE(w,\underline{x},\overline{x},\underline{y},\overline{y}) = 
    \begin{cases}
      w\ge \underline{x}y+x\underline{y}-\underline{x}\underline{y} \\
      w\ge \overline{x}y+x\overline{y}-\overline{x}\overline{y} \\
      w\le \underline{x}y+x\overline{y}-\overline{x}\underline{y} \\
      w\le \overline{x}y+x\underline{y}-\underline{x}\overline{y}
    \end{cases}       
\end{equation}
We introduce auxiliary variables for each of the four bilinear terms $\{a_i^{\phi},b_i^{\phi},c_i^{\phi},d_i^{\phi}\} = \{V_i^{\phi, R} I_i^{\phi, R},V_i^{\phi, I} I_i^{\phi, I},V_i^{\phi, R} I_i^{\phi, I},V_i^{\phi, I} I_i^{\phi, R}\}$ allowing us to convert constraints \cref{eq:bilinearP,eq:bilinearQ} to linear constraints with MCE constraints on each of the auxiliary variables. We also need additional constraints on the nodal current injections and nodal voltages in order to define the MCE constraints. These voltage and current bounds can be determined by applying a suitable preprocessing method using the nodal $P$ and $Q$ limits from the SMO bids. The resulting bounds will also implicitly satisfy constraints \cref{eq:Ibranchlims} and \cref{eq:Vmaglims} Thus, we can replace constraints \cref{eq:bilinearP,eq:bilinearQ,eq:Ibranchlims,eq:Vmaglims} with the following set of constraints in order to obtain the relaxed CI-OPF problem, which reduces to a linear program that can be solved easily. However, we do incur the overhead of computing the tightest possible $V$ and $I$ bounds to obtain a good convex relaxation, which in turn ensures that the relaxed solutions are feasible for the original problem.
\begin{subequations}
\label{eq:linear_mce}
\begin{align}
& P_i^\phi=a_i^{\phi} + b_i^{\phi}, \; \; Q_i^\phi=-c_i^{\phi} + d_i^{\phi} \quad \forall i \in \mathcal{N}, \phi \in \mathcal{P} \label{eq:linearPQ}\\
& \underline{I_{i}^{\phi,R}} \le I_{i}^{\phi,R} \le \overline{I_{i}^{\phi,R}}, \; \underline{I_{i}^{\phi,I}} \le I_{i}^{\phi,I} \le \overline{_{i}^{\phi,I}}  \label{eq:Ilims} \\
& \underline{V_{i}^{\phi,R}} \le V_{i}^{\phi,R}\le \overline{V_{i}^{\phi,R}}, \; \underline{V_{i}^{\phi,I}} \le V_{i}^{\phi,I}\le \overline{V_{i}^{\phi,I}} \label{eq:Vlims} \\
& a_i^{\phi} \in MCE(V_{i}^{\phi,R}I_{i}^{\phi,R},\underline{V_{i}^{\phi,R}},\overline{V_{i}^{\phi,R}},\underline{I_{i}^{\phi,R}},\overline{I_{i}^{\phi,R}}) \\
& b_{i}^{\phi} \in MCE(V_{i}^{\phi,I}I_{i}^{\phi,I},\underline{V_{i}^{\phi,I}},\overline{V_{i}^{\phi,I}},\underline{I_{i}^{\phi,I}},\overline{I_{i}^{\phi,I}}) \\
& c_{i}^{\phi} \in MCE(V_{i}^{\phi,R}I_{i}^{\phi,I},\underline{V_{i}^{\phi,R}},\overline{V_{i}^{\phi,R}},\underline{I_{i}^{\phi,I}},\overline{I_{i}^{\phi,I}}) \\
& d_{i}^{\phi} \in MCE(V_{i}^{\phi,I}I_{i}^{\phi,R},\underline{V_{i}^{\phi,I}},\overline{V_{i}^{\phi,I}},\underline{I_{i}^{\phi,R}},\overline{I_{i}^{\phi,R}}) \label{eq:mce_ineqs}
\end{align}
\end{subequations}

\subsubsection{Objective functions for voltage control}

After converting all quantities to per-unit (p.u.), we considered a weighted linear combination of several convex objective functions for the PM clearing using CI-OPF - where the weight $\xi$ controls the relative tradeoff between the first 2 `socio-economic' objectives versus the last 2 `electrical' objectives: 
\begin{align}
    & f^{obj}(x) = \sum_{\phi \in \mathcal{P}} \sum_{i \in \mathcal{N}_I}  \left[f_{i}^{\text{Load-Disutil},\phi}(x) + f_{i}^{\text{Gen-Cost}, \phi(x)} \right]  \nonumber \\
    & + \xi \sum_{\phi \in \mathcal{P}} \left[\sum_{(i,k) \in \mathcal{T}} f_{ik}^{\text{Loss},\phi}(x) + \sum_{i \in \mathcal{N}_I} f_{i}^{\text{Volt}, \phi}(x)\right] 
\end{align}
The first term minimizes disutility due to load flexibility:
$$
f_{i}^{\text{Load-Disutil},\phi}(x)= \beta_{i}^{P}(P_{i}^{L,\phi}-P_{i}^{L0,\phi})^2+\beta_{i}^{Q}(Q_{i}^{L,\phi}-Q_{i}^{L0,\phi})^2
$$
The second term minimizes generation costs. These are set by the locational marginal prices (LMPs) $\Lambda_i^P, \; \Lambda_i^Q$ for the primary feeder node at the point of common coupling (PCC) at the substation. For SMOs at all other primary feeder nodes, these depend on some fixed coefficients $\alpha_{i}^{P}, \; \alpha_{i}^{Q}$ that represent costs to the SMO for running its SM:
$$
f_{i}^{\text{Gen-Cost},\phi}(y) = \begin{cases} \Lambda_{i}^{P} P_{i}^{G,\phi}+\Lambda_{i}^{Q} Q_{i}^{G,\phi}, \text{if } i & \text{is PCC} \\
    \alpha_{i}^{P} P_{i}^{G,\phi} + \alpha_{i}^{Q} Q_{i}^{G,\phi} & \text{otherwise}
    \end{cases}
$$
The third term minimizes line losses in the network for more efficient operation. These are determined by the 
$$
f_{ij}^{\text{Loss},\phi}(x) = R_{ij}|I_{ij}^{\phi}|^2 = R_{ij}\left(I_{ij}^{{\phi,R}^2} + I_{ij}^{{\phi,I}^2}\right)
$$
where $\mathcal{T}$ is the set of network branches, $R_{ij}$ are branch resistances and $I_{ij}^{\phi}$ are branch current flows. These can be readily obtained from the nodal currents $I_i$ since $I = A^{\intercal} I_{branch}$, where $A$ is the three-phase graph incidence matrix.

The fourth term is the voltage regulation term that we specify to perform voltage control. This penalizes voltage deviations from some desired nominal values, in order to achieve a desired profile:
$$
f_{i}^{\text{Volt},\phi}(x) = \left(V_j^{\phi, R} - \tilde{V}_j^{\phi, R}\right)^2 + \left(V_j^{\phi, I} - \tilde{V}_j^{\phi, I}\right)^2
$$
In this study, we regulated voltage about setpoints $\tilde{V}_j^{\phi, R} = 1, \; \tilde{V}_j^{\phi, I} = 0$, to track a nominal magnitude $|\tilde{V}_j^{\phi}| = 1$ p.u. 

\subsection{Pricing}

Both the SM and PM result in localized, real-time prices for each DCA and SMO, respectively, which allows us to capture the high degree of spatial and temporal variation in prices. In this study, we focus on the pricing results for SMOs in the PM - please refer to \cite{nair2022hierarchical} for detailed results on localized retail tariffs for DCAs in the SM. We can derive PM prices by inspecting dual variables ($\lambda$) corresponding to different sets of linear equality constraints in the PM CI-OPF problem \cref{eq:ciopf}. The Lagrangian for the primal problem \cref{eq:ciopf} is:
\begin{align}
    \mathcal{L} & = f^{obj}(x) + \lambda_P^{\intercal} P_{balance} + \lambda_Q^{\intercal} Q_{balance} \nonumber \\
    & + \lambda_I^{\intercal} (I - YV) + \lambda_{ineq}^{\intercal} (RHS_{ineq} - LHS_{ineq})
\end{align}
where $P_{balance}$ and $Q_{balance}$ refer to the active and reactive power balance equations \cref{eq:bilinearP,eq:bilinearQ} respectively, and $I = YV$ enforces the linear Ohm's law constraint from \cref{eq:ohmslaw}. The last term in the Lagrangian corresponds to all the remaining inequality constraints from \cref{eq:Ilims}-\ref{eq:mce_ineqs}. However, our focus here is only on the duals of equality constraints \cref{eq:bilinearP,eq:bilinearQ,eq:ohmslaw} for pricing purposes. Note that the dual variable $\lambda_I$ is in terms of current, which can be converted to an equivalent value in terms of voltage:
\begin{align}
    & \lambda_I^{\intercal} (I - YV) \equiv \lambda_V^{\intercal} (ZI - V) = \lambda_V^{\intercal} (Y^{-1}I - Y^{-1}YV) \\
    & = \lambda_V^{\intercal}  Y^{-1} (I - YV) \implies \lambda_I^{\intercal} = \lambda_V^{\intercal} Y^{-1} \nonumber \implies \lambda_V = Y^{\intercal}  \lambda_I \nonumber
\end{align}
where $Z = Y^{-1}$ is the 3-phase network impedance matrix. These dual variables can be interpreted as prices for different services in the distribution grid. Thus, we propose the vector of dual variables above $\bm{\lambda} = [\lambda_P,\lambda_Q,\overline{\lambda_V}]$ as the distribution locational marginal price (dLMP) where $\overline{\lambda_V} = \mathrm{Re}(\lambda_V)$ is the real part of the complex dual variable. In particular, $\lambda_P$ and $\lambda_Q$ represent the P and Q-dLMP components for active and reactive power. The P-dLMP or energy price $\lambda_P$ is similar to the notion of locational marginal prices in the transmission system and WEM. Such a structure of P and Q components in a dLMP has also been proposed in \cite{bai2017distribution}, but we introduce the voltage support price $\overline{\lambda_V}$ in this paper for the first time. These dLMPs represent the overall grid services from DERs by providing real power, reactive power, and voltage support. We note that $\overline{\lambda_V}$ can be interpreted as a price for voltage control or regulation, because it reflects the effects of perturbations in the Ohm's law constraint, on our objective function, as shown below:
\begin{align*}
& \frac{\partial \mathcal{L}}{\partial V} = \frac{\partial f^{obj}(x)}{\partial V} - \lambda_I^{\intercal} Y = \frac{\partial f^{obj}(x)}{\partial V} - \lambda_V \\
& \text{At optimality} \; \frac{\partial \mathcal{L}}{\partial V^*} = \frac{\partial f^{obj}}{\partial V^*} - \lambda_V = 0 \implies \frac{\partial f^{obj}}{\partial V^*} = \lambda_V^*
\end{align*}
Thus, $\overline{\lambda_V}$ intuitively represents the costs of satisfying voltage constraints on the distribution grid (in terms of degrading the objective) and can be interpreted as the value of this voltage control grid service. Similarly, $\lambda_P$ and $\lambda_Q$ are costs associated with meeting power balance.

\section{Results and discussion}

\subsection{Numerical simulations}

We conducted a co-simulation of both the SM and PM on a modified IEEE-123 node feeder with high DER penetration comprising of rooftop solar photovoltaic (PV) systems, batteries, and flexible loads. The specifications of the modified network are shown in \cref{tab:specs}. The network was simulated using GridLAB-D in order to obtain realistic profiles for baseline power injections of SMOs and DCAs, as well as primary-level nodal voltages. Synthetic flexibility bids were then generated by randomly assigning flexibilities between 10-30\% for each of the DCAs. Simulations were conducted for a 24 h period, using weather data from San Francisco, CA on August 2, 2022, along with 5-min LMP data from the California Independent System Operator (CAISO). The SM was cleared every 1 min, while the PM was cleared every 5 min, in lockstep with the WEM.

\begin{table}
\centering
\begin{tabular}{ccc}
\toprule
 Type           &  Number        &  Capacity                       \\ \midrule
 DERs           &  380           &  1,745.8 kVA ($\approx$44\%)    \\
 PVs            &  207           &  880.84 kVA                     \\
 Batteries      &  173           &  865 kVA                        \\
 Spot loads     &  85            &  3,985.7 kVA                    \\
 Houses         &  1008          &  4-10 kW (variable)             \\
 Flexible loads &  1-2 per house &  10-50\% flexibility (variable) \\ \bottomrule
\end{tabular}
\label{tab:specs}
\caption{Specifications of modified IEEE 123-node feeder.}
\end{table}

\begin{figure}
\centering
     \includegraphics[width=\columnwidth]{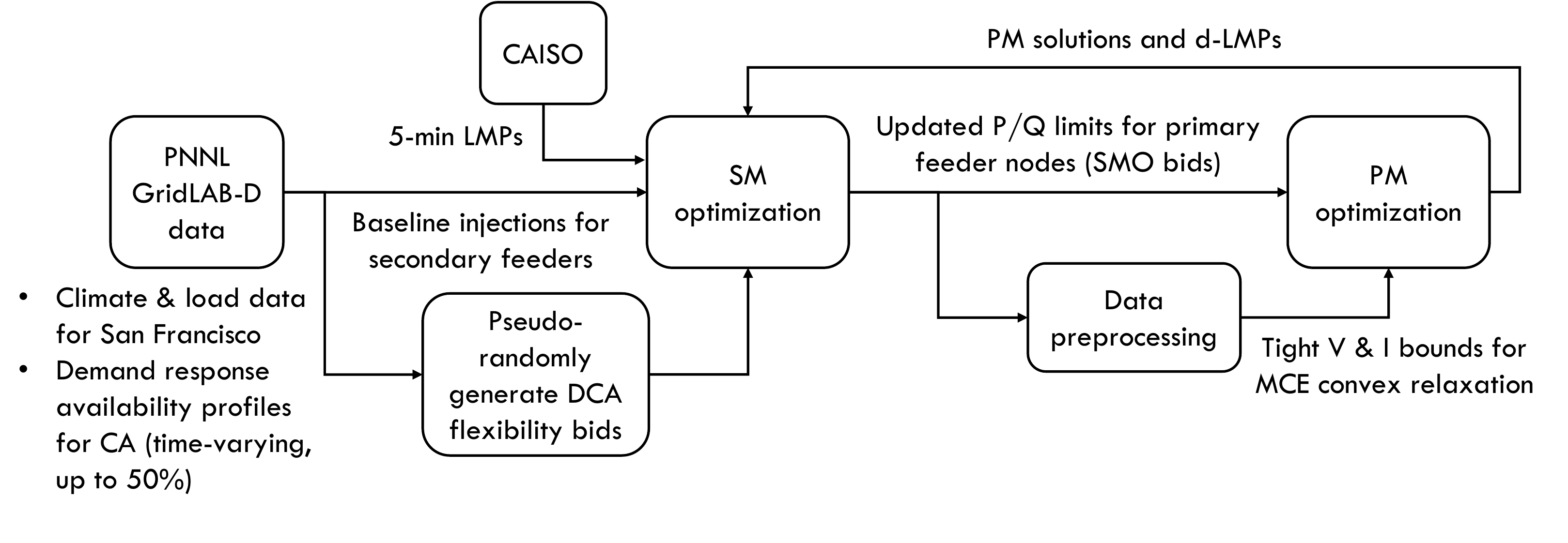}
     \caption{Workflow for SM and PM co-simulation. \label{fig:workflow}}
\end{figure}

The workflow for the co-simulation is shown in \cref{fig:workflow}. In particular, we feed in the aggregated solutions from the SM clearing to form the SMOs bids into the PM. These bids, which are in terms of active and reactive power flexibility ranges, are then preprocessed to give the corresponding V and I bounds needed for the MCE relaxation. The relaxed CI-OPF problem is then solved to clear the PM. We used the Gurobi solver for both the SM and PM optimization problem. In order to accelerate our simulations, we parallelized the SM clearing using MIT's Supercloud high-performance computing cluster \cite{reuther2018interactive} and Python's Message Passing Interface (MPI). At every secondary timestep $t_s$ (1 min), we solved the optimization problems for all 85 SMOs in parallel across multiple processors - making the problem much more computationally tractable and providing $\approx$ 80X speedup in solution runtimes. The dLMPs and nodal voltage solutions are 3-phase variables, but in the following sections, we calculate their mean values averaged over all the non-zero phases that are present at each node.

\subsection{Effects of the LEM on voltages}

\begin{figure}
     \centering
     \begin{subfigure}[t]{0.49\columnwidth}
         \includegraphics[width=\linewidth]{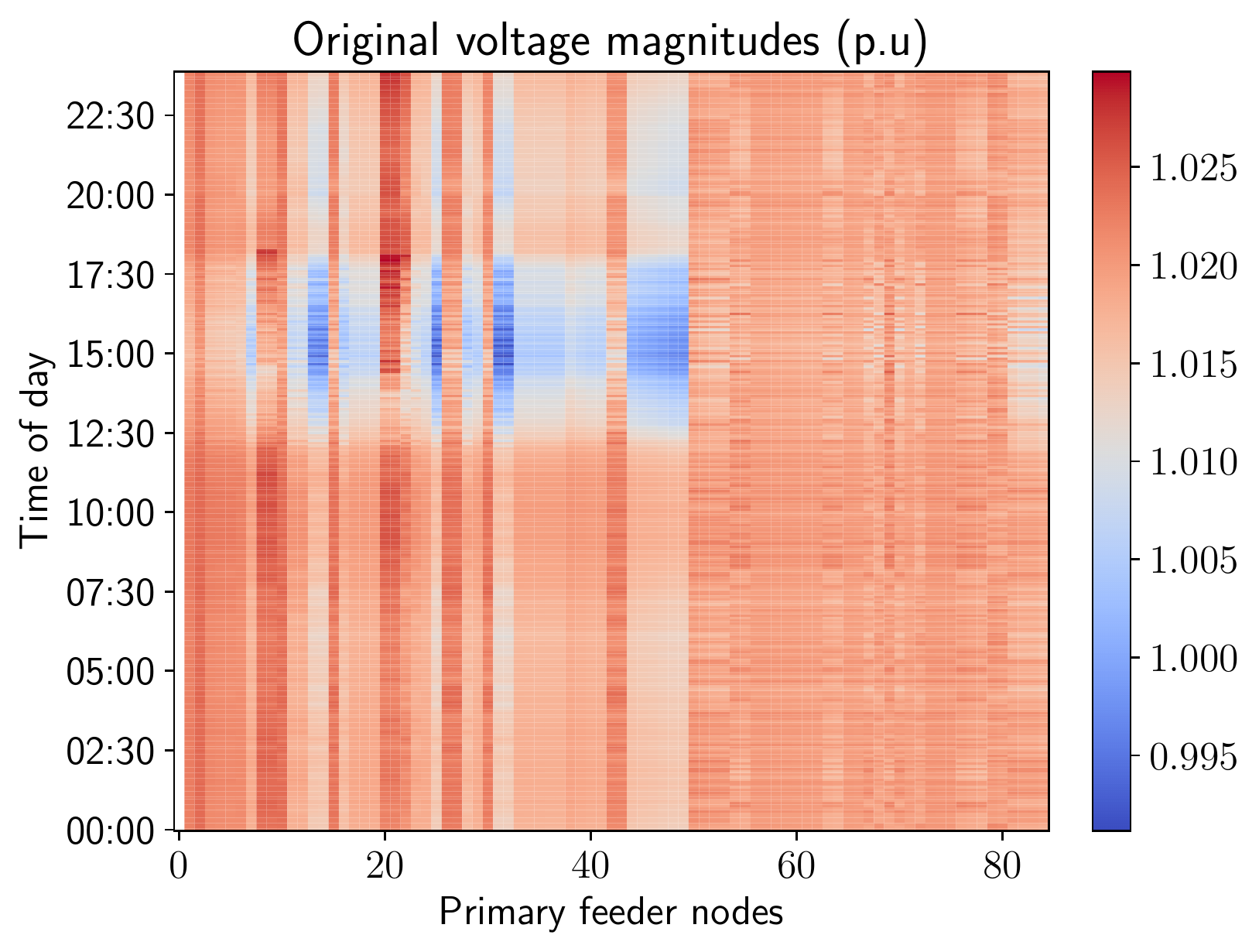}
         \caption{Without the LEM. \label{fig:Vmag_smos_orig}}
     \end{subfigure}
     \begin{subfigure}[t]{0.49\columnwidth}
         \includegraphics[width=\linewidth]{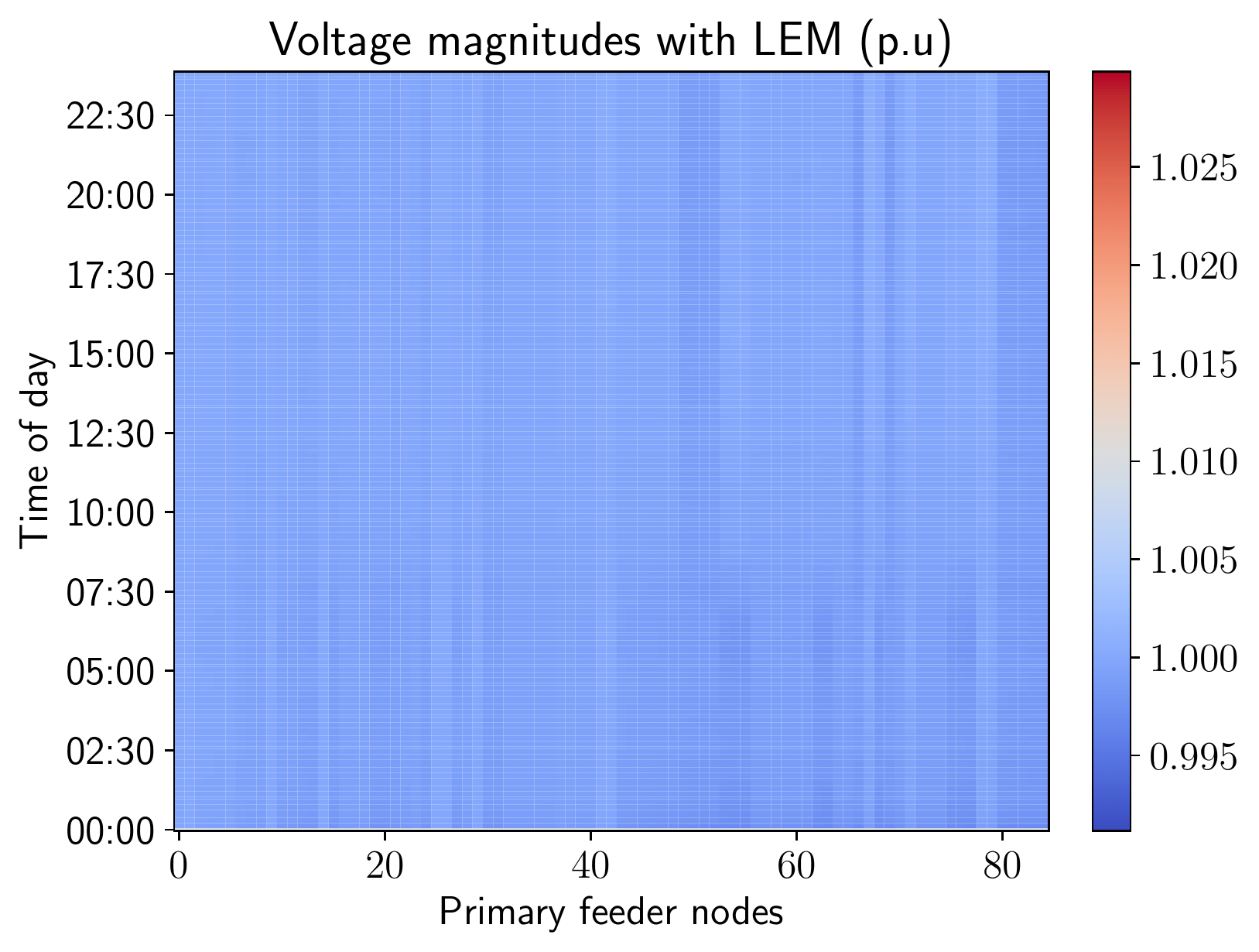}
         \caption{With the LEM. \label{fig:Vmag_smos_lem}}
     \end{subfigure}
     \caption{Primary level nodal voltage magnitudes with and without the LEM, at nodes with SMOs and over time.\label{fig:Vmag_effects}}
\end{figure}

We find that the LEM does indeed significantly improve the overall voltage profile by making it more uniform and bringing the voltage magnitudes closer to the desired 1 p.u. setpoint, as seen in \cref{fig:Vmag_effects}. In \cref{fig:Vmag_smos_orig} we notice overvoltage (i.e. $|V| >$ 1 p.u.) issues throughout most of the 24 h simulation period, but these are generally more pronounced during daylight periods of the day with higher PV output. Overvoltage problems are also more frequent and severe for specific primary nodes that correspond to SMOs and DCAs with greater local generation capacity from solar PV and/or batteries. Undervoltages (i.e. $|V| <$ 1 p.u.) are less common and occur during the afternoons, likely due to higher demand spikes from heating, ventilation, and cooling (HVAC) loads. The LEM is able to effectively coordinate DERs in order to mitigate both under and overvoltage issues throughout the day and across all nodes in the primary feeder, as seen in \cref{fig:Vmag_smos_lem}. This is achieved through smarter scheduling and dispatch of resources - these actions may include (but are not limited to) controlled battery charging or discharging, power factor control using smart inverters as well as shifting or curtailment of flexible loads and appliances. This results in more uniform spatial and temporal voltage distributions.

\begin{figure}
     \centering
     \begin{subfigure}[t]{0.49\columnwidth}
         \includegraphics[width=\linewidth]{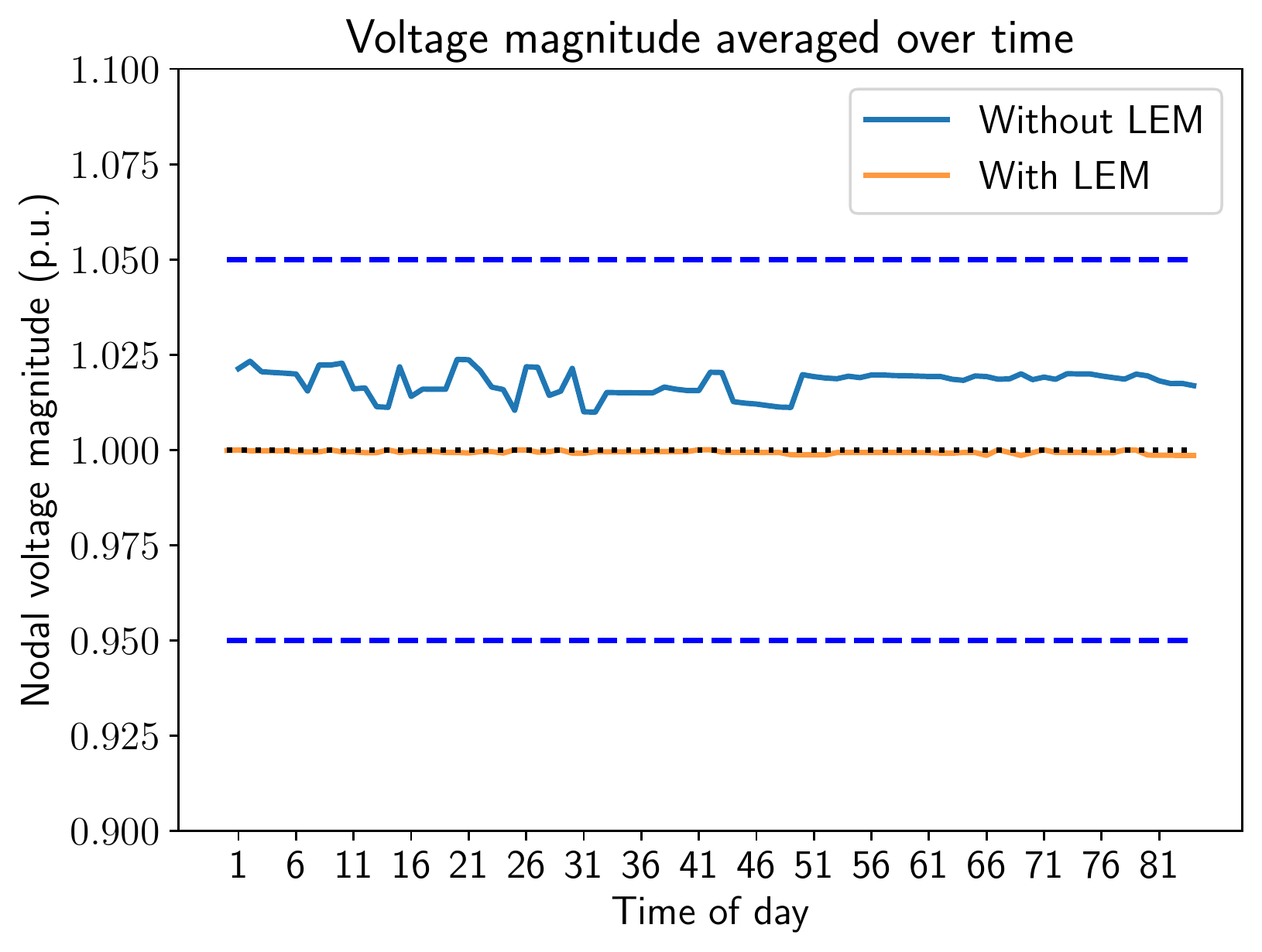}
        \caption{Nodal averages over time. \label{fig:Vmag_avg_overnodes_time}}
     \end{subfigure}
     \begin{subfigure}[t]{0.49\columnwidth}
         \includegraphics[width=\linewidth]{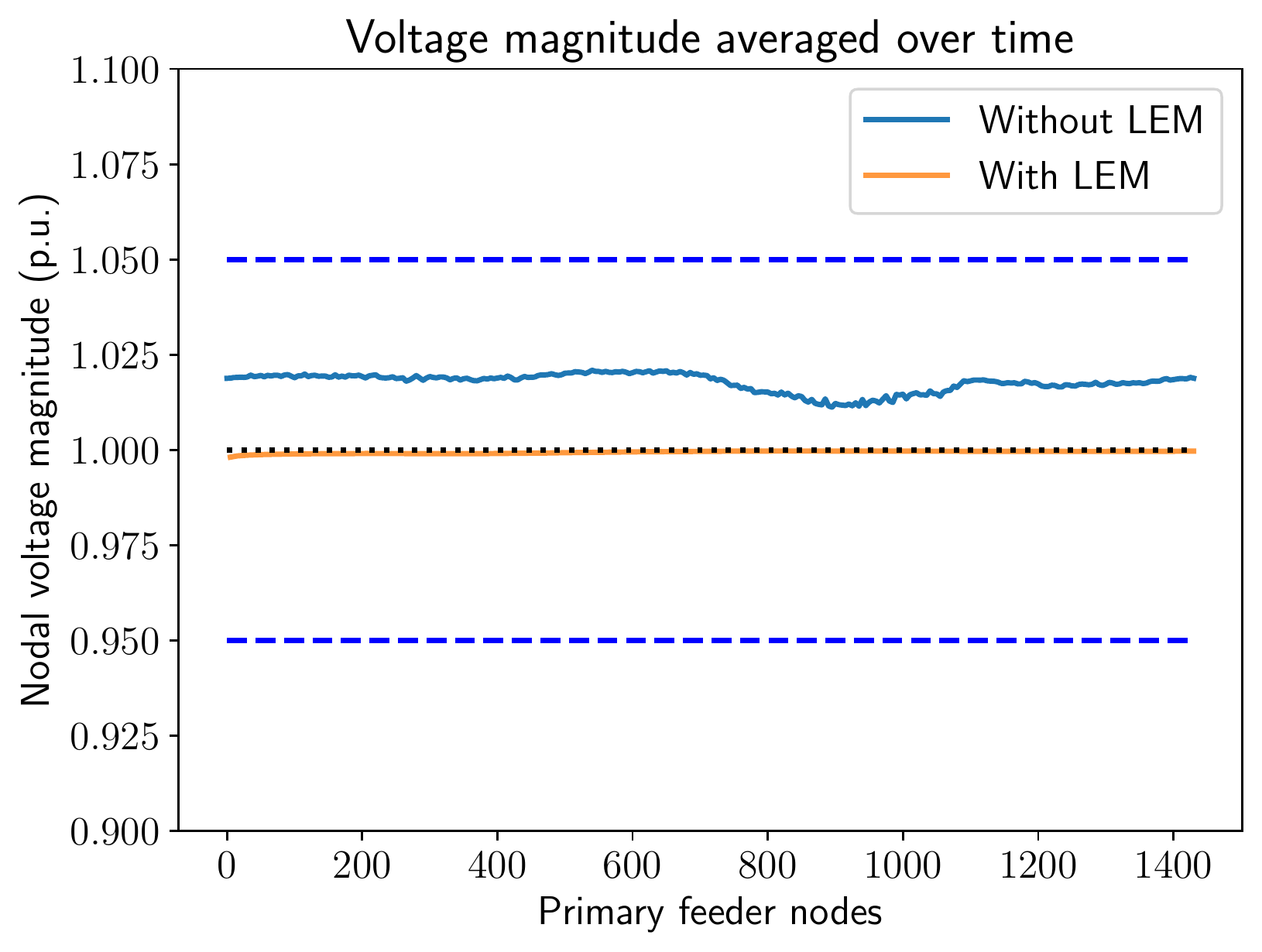}
        \caption{Daily average across nodes. \label{fig:Vmag_avg_overtime_nodes}}
     \end{subfigure}
     \caption{Primary level nodal voltage magnitudes with and without the LEM, at nodes with SMOs and over time. \label{fig:Vmag_avgs}}
\end{figure}

The voltage profile improvements are also evident from \cref{fig:Vmag_avgs}, where both the spatial (in \cref{fig:Vmag_avg_overnodes_time}) and temporal (in \cref{fig:Vmag_avg_overtime_nodes}) mean voltage magnitudes are almost exactly equal to the desired 1 p.u. with the LEM in place, as opposed to the consistently higher mean voltages observed without the LEM. The voltages are also well within the ANSI safe operating voltage limits of $[0.95,1.05] \; p.u.$. 


\subsection{dLMP results}

\begin{figure}
     \centering
     \begin{subfigure}[t]{0.49\columnwidth}
         \includegraphics[width=\linewidth]{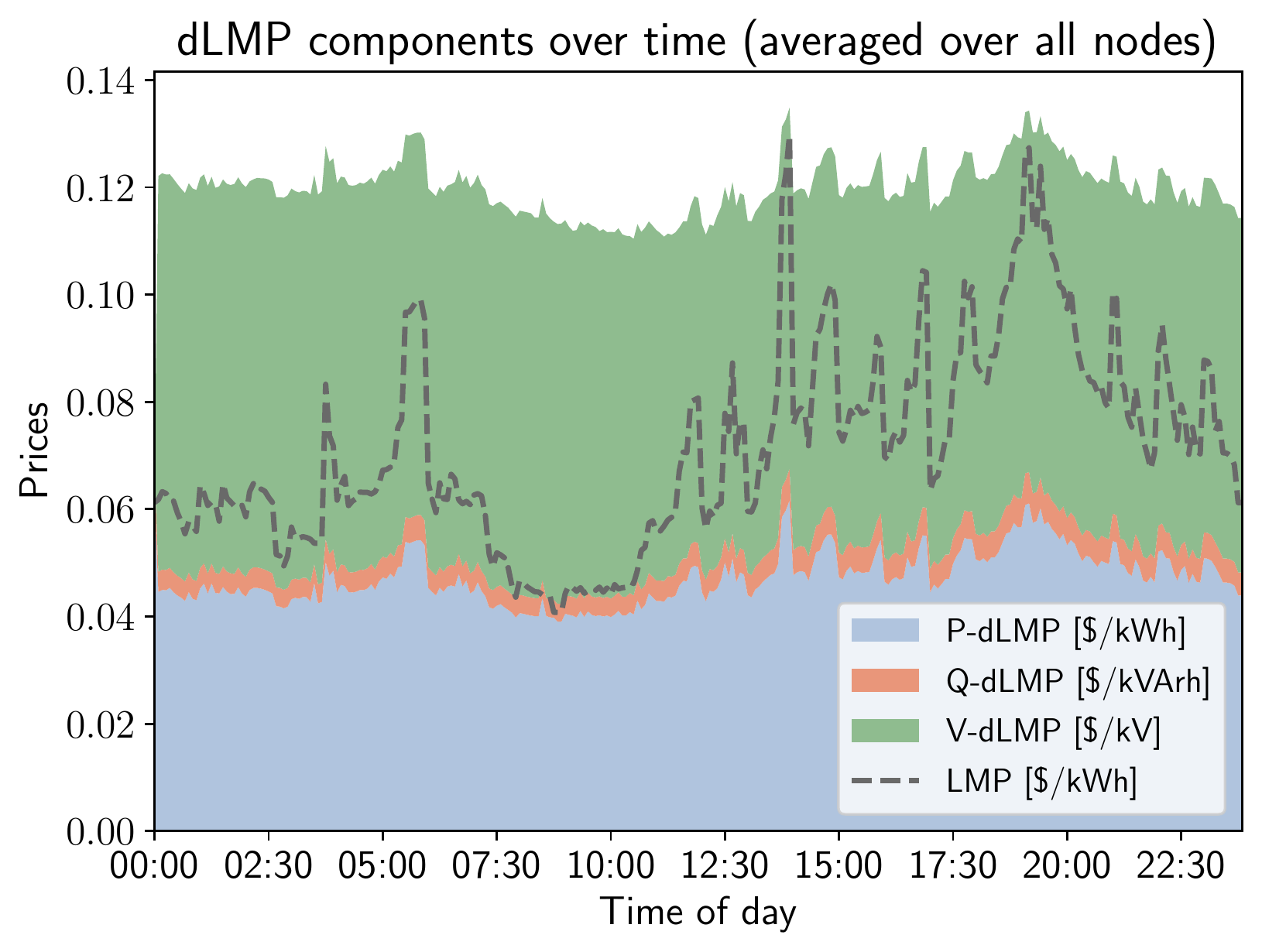}
         \caption{dLMP components over time, averaged over all primary level nodes along with the LMP. \label{fig:mu_avg_nodes}}
     \end{subfigure}
     \begin{subfigure}[t]{0.49\columnwidth}
         \includegraphics[width=\linewidth]{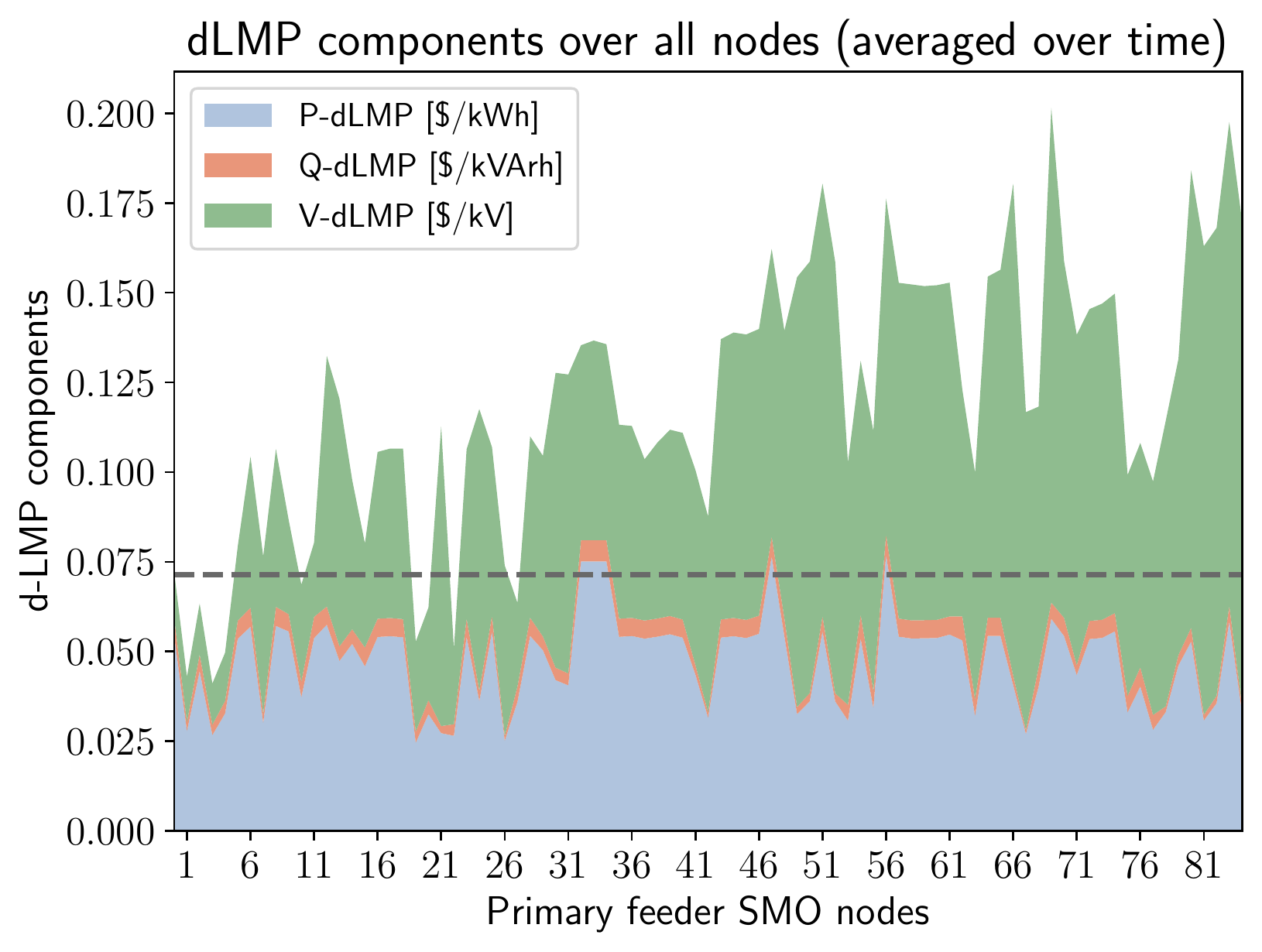}
         \caption{dLMP components over primary level nodes, averaged over the entire 24 h simulation period. \label{fig:mu_avg_time}}
     \end{subfigure}
     \caption{Variations in dLMPs for over nodes and time. \label{fig:muavg_trends}}
\end{figure}

Fig. \ref{fig:muavg_trends} summarizes the PM pricing results and decomposition of the dLMPs into the three components of P, Q, and V support prices. In \cref{fig:mu_avg_nodes}, temporal variations of the dLMP components are shown over the whole day, when averaged over all the SMO nodes. At all times, the mean dLMP over the primary feeder is higher than the LMP at the substation or PCC. This makes intuitive sense since the dLMP accounts for additional costs and losses in the distribution grid downstream of the transmission grid, that are not included in the LMP. This also allows the DSO and PMOs to recoup their own costs for running the retail markets while participating in the WEM. Another interesting result is that throughout the day, the P and V-dLMP components contribute to the bulk of the dLMP, while the Q-dLMP only makes up a small portion of the price. This makes sense since nodal Q injections are much smaller in magnitude compared to P injections across the distribution feeder, and is also in line with other works that have suggested for instance, that Q-dLMPs should roughly be $\approx \; 10\%$ of the corresponding P-dLMPs \cite{federal_energy_regulatory_commission_payment_2014}. Another reason for the small Q price contribution could be that reactive compensation plays a key role in maintaining grid voltages, so some of its effects may already be taken into account by the V-dLMP.


\begin{figure}
     \centering
     \begin{subfigure}[t]{0.49\columnwidth}
         \includegraphics[width=\linewidth]{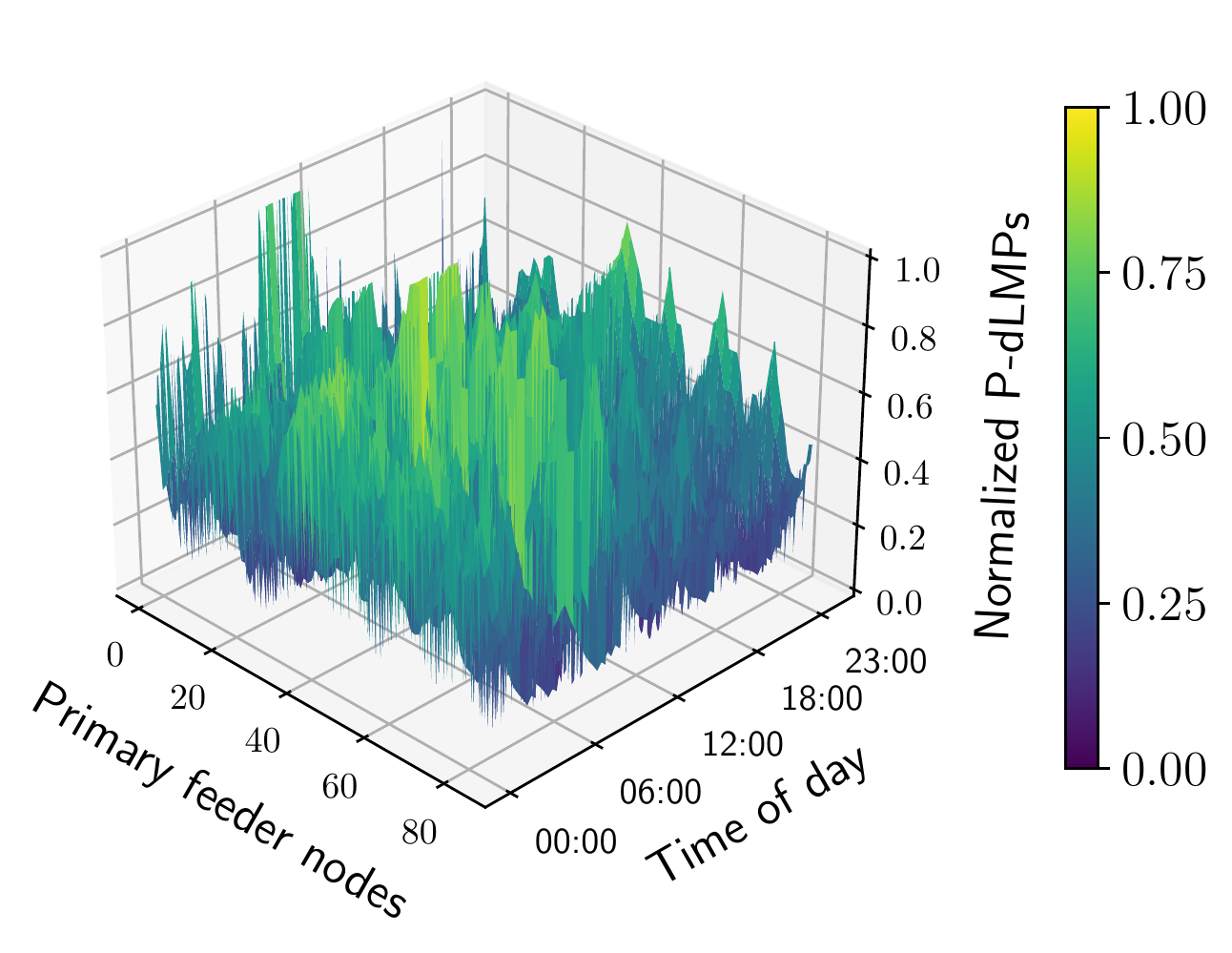}
         \caption{P-dLMP \label{fig:muP_dist}}
     \end{subfigure}
     \begin{subfigure}[t]{0.49\columnwidth}
         \includegraphics[width=\linewidth]{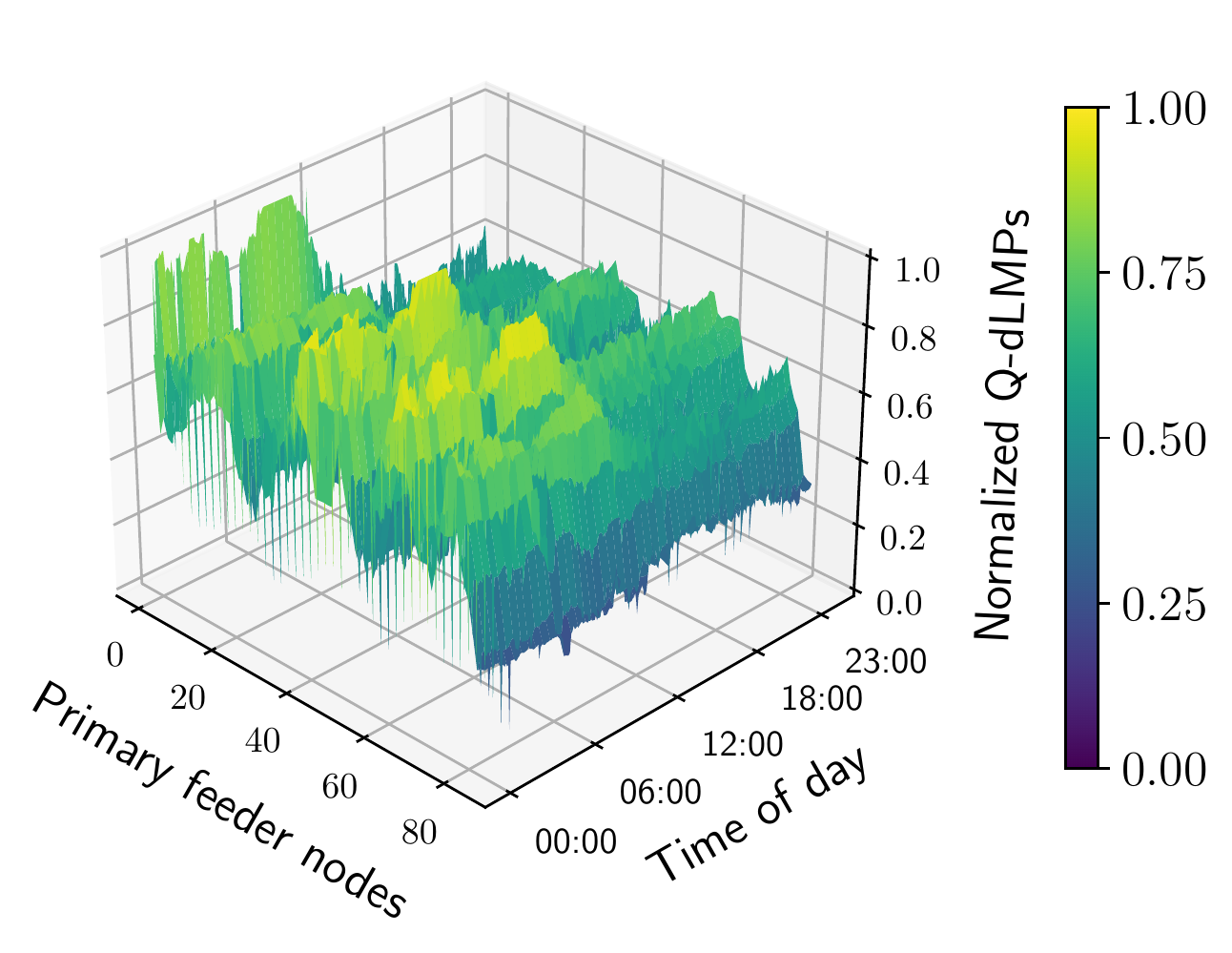}
         \caption{Q-dLMP \label{fig:muQ_dist}}
     \end{subfigure}
      \begin{subfigure}[t]{0.49\columnwidth}
     \includegraphics[width=\linewidth]{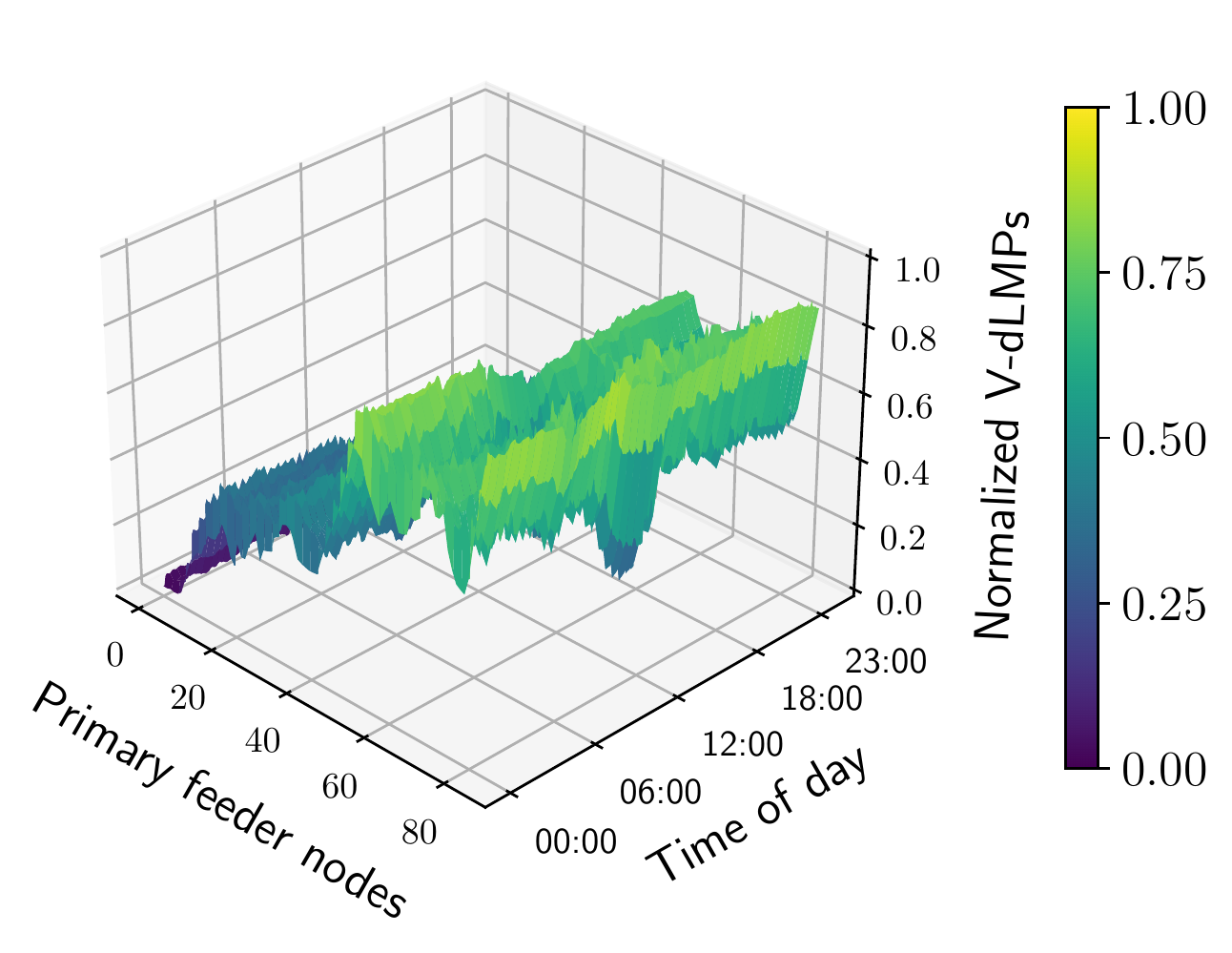}
     \caption{V-dLMP \label{fig:muV_dist}}
    \end{subfigure}
     \caption{Distributions of dLMP components over all SMO nodes and over the 24 h simulation period. \label{fig:mu_dist}}
\end{figure}

In \cref{fig:mu_avg_time}, the spatial node-to-node variations of the time-averaged dLMPs are shown, along with the average LMP for the day. We see again that the combined average P, Q, and V-dLMPs are higher than the average LMP at most nodes, except for a few of them ($< \; 10$). The relative breakdowns of P versus Q-dLMPs are roughly similar across the network, but the contributions of the V-dLMP differ quite significantly for different nodes. For e.g., the V-dLMP is relatively much larger for node 71 in \cref{fig:mu_avg_time}, indicating that it may be more challenging to meet grid physics constraints and support voltages at these specific nodes, while solving the PM clearing and CI-OPF problem. Further analysis is necessary to fully interpret and explain this trend, this will be explored more as part of future work. Both plots in \cref{fig:muavg_trends} also show that the costs associated with voltage support are significant and must also be adequately accounted for in retail markets, rather than focusing solely on P and Q energy prices. In both \cref{fig:Vmag_avg_overnodes_time,fig:Vmag_avg_overtime_nodes}, the combined P and Q-dLMPs without including the V-dLMP are consistently lower than the LMP. This is in agreement with other related works such as \cite{bai2017distribution,haider2021reinventing} - this indicates that distribution level costs involve not just those associated with satisfying power balance, but also other constraints like Ohm's law (\cref{eq:ohmslaw}).


The locational-temporal variations of the normalized P, Q, and V-dLMPs are shown in \cref{fig:mu_dist}, for all 85 primary feeder nodes with SMOs and over the 24 h period. We observe that there's a great deal of variability in these prices, which further motivates the crucial need for new retail market structures such as our LEM in order to capture these variations. This would allow us to accurately compensate different resources depending both on the time of day as well as their geographic locations within the distribution system. Another important observation is that our combined dLMP is significantly lower than the current retail rate charged by utilities and other load-serving entities (LSE), throughout the day and across all primary nodes. Since current retail rates only include active power, we calculate an equivalent rate $\lambda_{eq}$ in \$/kWh for our LEM as a weighted average of all 3 dLMP components:
\begin{align}
    \lambda_{eq} & = (\lambda_P^* P^* + \lambda_Q^* Q^* + \overline{\lambda_V}^* \Delta V^*)/P^* \nonumber \\
    \Delta V^* & = |V^{R^*} - 1| + |V^{I^*}|
\end{align}
where $\Delta V^*$ are the deviations of voltages from the nominal values. The average bundled tariff for Pacific Gas \& Electric (PG\&E) customers in August 2022 was $33.72 \; \mbox{\textcent}/kWh$, compared to the mean equivalent rate $\overline{\lambda_{eq}} = 5.38 \; \mbox{\textcent}/kWh$ in our LEM, averaged over the day and the whole network. This represents a $\approx \; 84 \%$ reduction, indicating the LEM is able to coordinate and schedule DERs more effectively to reduce network-wide costs. These tariffs are likely to increase further as higher DER penetration places more stress on distribution grids, but our LEM can help mitigate these challenges \cite{2019OperationalCommittee}.

However, it should also be noted that in this paper, we have only included costs for operating the primary market while meeting power flow constraints imposed by grid physics. In reality, the DSO incurs additional costs such as maintenance costs, infrastructure expenses, and delivery charges as well as profit margins imposed by LSEs. In addition, it has to recoup its costs for importing power from the WEM and transmission grid. For similar reasons, the retail rates charged by the SMOs to its DCAs may be higher than the breakeven tariffs determined by \cref{eq:breakeven_tariff}. These additional costs may reduce the reported margin of improvement from 64\% to a certain extent. The final dLMPs and retail rates also represent the value provided by the PMOs and SMOs (as well as the DSO that oversees both) in terms of serving demand as well as by facilitating market participation for DERs. They allow DERs to actively bid into retail and wholesale markets and get appropriately compensated for services they provide to the grid. This also motivates recent regulations like FERC order 2222 which opened up WEM participation to DERs \cite{FERCCommission}, as well as the push towards performance-based rate regulation - which evaluates actual utility performance when establishing rates as an alternative to calculating rate plans based on utility capital investments \cite{Performance-BasedPBR}.


\section{Conclusion and future work}

In this paper, we applied a hierarchical LEM to provide distribution grid services. We solved ACOPF at the primary level using a current injection model valid for balanced and meshed networks. We also accurately decomposed prices for such services, by deriving dLMP components corresponding to P, Q, and voltage support. Numerical simulations show that the LEM successfully mitigates under and overvoltage issues across the network throughout the day. We found V-dLMP does contribute significantly relative to the P and Q-dLMPs, and the costs associated with meeting critical grid physics constraints must be taken into account for pricing. The LEM also captures the high spatial and temporal price variations, which enables setting time-varying, differentiated local tariffs for each node. Finally, by optimally coordinating DERs, the LEM achieves much lower costs for grid operators and lower rates for customers, relative to the status quo. For future work, we will compare our LEM rigorously against other proposed markets, extend to other grid services like conservation voltage reduction, and test performance on larger networks with more severe voltage issues.

\section*{ACKNOWLEDGMENTS}

The authors thank Rabab Haider, Giulio Ferro and Venkatesh Venkataramanan for valuable discussions as well as the MIT SuperCloud and Lincoln Laboratory Supercomputing Center for computational resources that have contributed to the research results reported within this paper. 

\bibliographystyle{IEEEtran}
\bibliography{refs_manual}

\end{document}